\documentclass[%
 aip,
 sd,%
 amsmath,amssymb,
 reprint,%
]{revtex4-1}
\usepackage{subfigure}
\usepackage{graphicx}
\usepackage{dcolumn}
\usepackage{bm}

\begin{document}

\preprint{AIP/123-QED}

\title[]{Quantum hydrodynamics in the rotating   reference frame}

\author{Mariya Iv. Trukhanova}
 \altaffiliation[ ]{Faculty of physics, Lomonosov Moscow State University, Moscow, Russian Federation}
 \email{mar-tiv@yandex.ru}

\date{\today}

\begin{abstract}
    In this paper we  apply quantum hydrodynamics (QHD) to  study   the quantum
evolution of a system of spinning  particles and particles that have the electric dipole moments EDM  in the rotating  reference frame. The  method presented
is based on the many-particle microscopic  Schrodinger equation in the rotating   reference frame. Fundamental QHD equations for charged or neutral spinning and EDM-bearing  particles
were shaped due to this method and    contain  the spin-dependent inertial force field.  The polarization dynamics in systems of neutral  particles in the rotating frame is shown to cause formation of
a new type of waves, the dipole-inertial waves. We have analyzed  elementary excitations in a system of neutral polarized fluids placed into an external electric field in 2D and 3D cases.
We predict the novel type of 2D dipole-inertial wave and 3D - polarization wave modified by rotation in systems of  particles with dipole-dipole interactions.
\end{abstract}

\maketitle

\begin{quotation}

\end{quotation}

 \section{Introduction}

Over many  years  scientists  paid attention to the effects occurring in the non-inertial frames.  The influence of the inertial effects on electrons  have always been  the main  focus of
most studies,  starting with those that belong to   Barnett  and Einstein - de
Haas   \cite{2}, \cite{3}. Recently the
 influence of  rotation in spintronic applications has been noticed.  The spin-dependent inertial
 force in an accelerating system under the presence of electromagnetic
fields have been derived from the generally covariant Dirac equation \cite{1}.
It was shown that mechanical vibration in a high frequency resonator can create a spin current via
the spin-orbit interaction augmented by the linear acceleration.    The enhancement of the spin-rotation coupling due to the interband mixing was predicted in Ref. \cite{110}. The theoretical investigation of the inertial effects' influence  on the spin dependent transport of
conduction electrons in a semiconductor  was reviewed  in Ref. \cite{4}.
Equations of motion for a
wavepacket of electrons in the two-dimensional planes subject to the spin-orbit interaction were derived
in \cite{5}, using the inertial effects due to the mechanical rotation. The author obtained
 a superposition
of two cyclotron motions with different frequencies and a circular spin current.

A homogeneous fluid rotating with constant angular velocity
leads to the emergence of  an  unusual class of inertial waves spreading
 in the interior of the fluid or
inertial waves \cite{7} - \cite{11}.
Inertial waves, caused by fluid rotation, are circularly polarized waves with a sense of rotation fixed by their helicity.
Inertial waves
 have many important
features and play an important role in   Geophysics \cite{11}, \cite{13}, in  evolution of  liquid planet core \cite{14}, \cite{15},
rotating stars \cite{16}.  The restoring force for such kind of waves in the  rotating frame of reference  is the Coriolis force, which
leads  to the circular motions of
fluid elements.
Large-scale geophysical and astrophysical flows are populated by a great variety of internal
waves, some maintained by density stratification (internal gravity waves), some by the
background planetary or stellar rotation (inertial waves), and yet others by the large-scale
magnetic fields which thread through interplanetary space and are generated in the interiors
of planets and stars (Alfv´en waves) \cite{8}.

The investigation of inertial waves had been implemented in various geometries. The viscous flow inside
a closed rotating cylinder of gas subjected to periodic axial compression had been investigated numerically \cite{161}.
An experimental study of inertial waves in a closed cone had been  presented \cite{162}, in which the inertial  waves  are excited  by
a  slight  periodic  oscillation  superimposed  on  the cone's basic rotation rate.

The experimental visualization of inertial waves had been realized using particle image
velocimetry \cite{17}. The author presented direct visualization of the velocity and vorticity fields in a
plane normal to the rotation axis and determined  the characteristic
wavelength.  The dynamics of the anisotropy of grid-generated decaying turbulence in a rotating frame had been experimentally investigated
by means of particle image velocimetry on the large-scale "Coriolis" platform  \cite{18}, \cite{19}.

The dipole-dipole interactions  are the longest-range interactions possible between two neutral atoms or molecules and occur in  many
             systems in nature.   This interaction is one of the most important interactions between atoms or
molecules.   In recent years much attention is paid to the effect of the intrinsic electric
dipole moment (EDM) on the characteristics of
charged and neutral particle systems. The propagation of perturbation
of  EDM does not require much energy as it occurs
without mass transfer. This process may be used in  the information transfer. In biological systems, for example,
polarization processes, i.e. EDM propagation, are the
predominant way of signal transfer \cite{E0}, \cite{E00}.

Dipole-dipole interactions,  are studied and manipulated in Rydberg atoms, provide a strong coupling
between atoms in an ultracold Rydberg gas,  because these atoms have high principal quantum number and  have large electric transition dipole moments compared to ground state
atoms \cite{E6}, \cite{E7}.  This interaction
can be tuned into resonance with a small electric field. This feature  of Rydberg atoms can be used in quantum computing \cite{E71}, in the  quantum
information processing with cold neutral atoms \cite{E72}. Generation of entanglement between two individual Rydberg  $Rb^{87}$ atoms in hyperfine ground states had been observed in \cite{E71} using Rydberg blockade effect.    Dipole-dipole  interactions
between Rydberg atoms had been first observed by Raimond
 using spectral line broadening \cite{E8}. In the magneto-optical trap ground-state atoms are cooled
down to $100 µK$ by  laser cooling \cite{E81}.  Recently it has
been shown that Rydberg excitation densities in a
magneto-optical trap  are limited by these interactions \cite{E9}. Another way is deceleration
and trapping of Rydberg atoms by static electric fields \cite{E91}.
 Resonant electric dipole-dipole interactions between cold Rydberg atoms had been observed using the
microwave spectroscopy \cite{E10}. The attraction of Rydberg atoms is that it is possible to tune the Rydberg energy levels through resonance for the dipole$–$dipole energy transfer. The investigation of  the properties of a cold ( $~100µK$) and dense ( $\sim 10^{10}cm^{-3}$) atomic Rydberg Cs gas had been studied in \cite{E11} using   a $"$frozen Rydberg gas$"$ model.

Classical methods were used previously to create a description
of the collective dynamics of  particles in the rotating frame
that takes into account the inertial effects \cite{7} - \cite{10}. We will use a  quantum mechanics description for systems of N
interacting particles is based upon the many-particle Schrodinger equation (MPSE)  that specifies a wave function
in a 3N-dimensional configuration space of inertial frame. The many-particle quantum hydrodynamics in the inertial frame was investigated in Refs. \cite{190} - \cite{230}. As wave
processes, processes of information transfer, diffusion and
other transport processes occur in the three-dimensional
physical space of rotating frame, a need arises to turn to a mathematical
method of physically observable values which are determined
in a 3D physical space. To do so we should derive
equations those determine dynamics of functions of
three variables, starting from MPSE in the rotating frame. This problem has
been solved with the creation of a method of many-particle
quantum hydrodynamics (MPQHD) in the non-inertial frame.

\section{Construction  of fundamental  equations
and the model accepted}

In this section we derive the many-particle
quantum hydrodynamics ({\em MPQHG}) equations from
many-particle Pauli-Schrodinger equation ({\em MPSE}). We receive the equations for the system
of charged particles with spins. Method of MPQHG allows to present dynamic of system
of interacting quantum particles in terms of functions defined
in 3D physical space. It is important at investigation
of wave process, which take place in 3D physical space.

For the beginning we present the one-particle Hamiltonian in the rotating frame which can be derived from the fundamental equation
for spinning particle in a curve space-time  \cite{1}

        \begin{equation} \label{Dirac}   \Biggl( \gamma^{\mu} \Biggl[\partial_{\mu}-\Gamma_{\mu}-
        \frac{iqA_{\mu}}{\hbar}   \Biggr]
        +\frac{mc}{\hbar}  \Biggr)\psi=0,\end{equation}
 where $c, \hbar$ and $q$ are the speed of light, the Planck constant and charge of an electron respectively,
 $\Gamma_{\mu}$ is the spin connection \cite{1}. The $4-spinor$ wave function $\psi$ contains
 the spin-up and spin-down  components. The Hamiltonian for the system
of charged particles contains the  terms of the induced electric and magnetic
fields due to inertial effect and can be obtained from the one-particle Dirac equation (\ref{Dirac}). The many-particle  Hamiltonian in the rotating frame can be obtained as

  \begin{equation}\label{H} \hat{H}=\hat{H}_{0}+\hat{H}_{rotor}+\hat{H}_{SO}\end{equation}

  $$  \hat{H}_0=\sum^{N}_{p=1}\Biggl({\frac{\hat{D}^{2}_{p}}{2m_p}+q_p\varphi_{p,ext}-\mu_p\hat{\sigma}^{\alpha}_{p}B^{\alpha}_{p,ext}}\Biggr)
$$
\begin{equation} \label{H1}\qquad\qquad+\frac{1}{2}\sum^{N}_{p\neq n} q_pq_nG_{pn}-\frac{1}{2}\sum^{N}_{p\neq n,n}\mu^{2}_{p}F^{\alpha\beta}_{pn}\hat{\sigma}^{\alpha}_{p}\hat{\sigma}^{\beta}_{n},   \end{equation}

\begin{equation} \label{H2}
 \hat{H}_{rotor}= -\sum^{N}_{p=1}\Biggl(\varepsilon^{\alpha\beta\gamma}\Omega^{\alpha}_p\cdot r^{\beta}_p\hat{D}^{\gamma}_p +\frac{\hbar}{2}\hat{\sigma}^{\alpha}_{p}\cdot\Omega^{\alpha}_{p}\Biggr), \end{equation}

  $$  \hat{H}_{SO}=-\sum^{N}_{p=1}\frac{\mu_p}{m_pc}\varepsilon^{\alpha\beta\gamma}\hat{\sigma}^{\alpha}_{p}\cdot E^{\beta}_{p,ext}\hat{D}^{\gamma}_p  $$
 $$ +\sum^{N}_{p\neq n}\frac{q_p\mu_n}{m_nc}\varepsilon^{\alpha\beta\gamma}\hat{\sigma}^{\alpha}_{n}\cdot \partial^{\beta}_{p}G_{pn}\hat{D}^{\gamma}_n $$
     \begin{equation} \label{H3}                                              + \sum^{N}_{p=1}\frac{\mu_p}{m_pc}\varepsilon^{\alpha\beta\gamma}\varepsilon^{\beta\mu\nu}\varepsilon^{\nu ij}\hat{\sigma}^{\alpha}_{p}\cdot B^{\mu}_{p,ext}r^j_p\Omega^i_p\hat{D}^{\gamma}_p. \end{equation}

We consider a system of $N$ interacting particles. The
following designations are used in the Hamiltonian (\ref{H}): $\hat{\textbf{D}}_p=-i\hbar\hat{\nabla}_p-\frac{q_p}{c}\textbf{A}_p,$
where     $\textbf{A}_{ext}, \varphi_{p,ext} $  -  are   the vector and scalar potentials of
external electromagnetic field, $\mu_p=g\mu_B/2$, $\mu_{p}$ - is the electron  magnetic moment
and $\mu_{pB}=q_p\hbar/2m_pc$  is the Bohr magneton,  $q_p$ stands for the charge of electrons $q_e=-e$ or for the charge of ions $q_p=e$, and $\hbar$ is
the Planck constant, $g\simeq 2.0023193$, $m_p$ denotes the mass  of  particles, c
 is the speed of light in vacuum. In this case the many-particle Pauli-Schrodinger equation for quantum
particles motion in the external fields in the rotating frame has the form

\begin{equation}
            i\hbar\frac{\partial\Psi(R,t)}{\partial t}=\hat{H}\Psi(R,t),
  \end{equation}
where $R=(\vec{r}_{1},...,\vec{r}_{N})$ and $\Psi(R,t)$ is the $rank-N spinor$.  The first term at the right-hand side of the (\ref{H1})
gives sum of kinetic energies of all particles, it
contains long derivative including action of the vortex part of the external electromagnetic
field on the particle charge. The second term in (\ref{H1}) describes potential energy of charges in the external
electric field. The third term describes the Zeeman effect with an external magnetic field and represents the Zeeman energy.
The fourth term in (\ref{H1}) presents the Coulomb interaction between particles and the last term  characterizes the spin-spin interactions.
Green functions of  the Coulomb and spin - spin interactions   are $ G_{pn}=1/r_{pn}, \qquad  F^{\alpha\beta}_{pn}=4\pi\delta_{\alpha\beta}\delta(\vec{r}_{pn})+\partial^{\alpha}_{p}\partial^{\beta}_{n}(1/r_{jk}).  $
The Hamiltonian $(\ref{H2})$  characterizes the inertial effects. The first term in (\ref{H2}) describes the mechanical rotation coupling with the angular momentum of particles $\textbf{r}_p\times\hat{\textbf{D}}_p$ and leads to the Euler,
Coriolis and centrifugal forces in the equation of motion. The second term is the spin-rotation coupling and results in  the Barnett effect.
The spin-rotation coupling can be unified with the Zeeman energy $\mu_p\mathbf{\hat{\sigma}}_{p}\cdot(\mathbf{B}_{p}+\mathbf{B}_{\Omega})$
using the effective Barnett field of the particle $\mathbf{B}_{\Omega}=cm_p\mathbf{\Omega}/q_p$.

The   spin-orbit coupling is characterized by the terms of Hamiltonian (\ref{H3}).
The first term on the right-hand side  represents the interaction of internal angular momentum or spin with the external electric field and
the second term describes the effect of the Coulomb electric field on the spin. The third term in the Hamiltonian (\ref{H3}) describes the influence
of the effective electric field on the moving spin in the inertial frame. The effective spin-dependent electric field  arises
in the rotating frame.

The first step in the construction of MQHD apparatus for the inertial frame is
to determine the concentration of particles in the neighborhood
of $\mathbf{r}$ in a physical space.  If we define the concentration
of particles as quantum average of the concentration
operator in the coordinate representation $\hat{n}=\sum^{N}_{p}\delta(\mathbf{r}-\mathbf{r}_{p})$
we arrive at the conclusive definition for the concentration

             \begin{equation} \label{n}
n(\mathbf{r},t)=\int dR\sum^{N}_{p}\delta(\mathbf{r}-\mathbf{r}_{p})\Psi^{+}(R,t)\Psi(R,t)=\langle\Psi^{+}\Psi\rangle.
\end{equation}

We will use the definition for the average value of the  physical quantity  $f(R,t)$  of the particles

\begin{equation} \label{a}
\langle f\rangle=\int dR\sum^{N}_{p}\delta(\mathbf{r}-\mathbf{r}_{p})f(R,t).
\end{equation}

Differentiation of $n(\mathbf{r},t)$ with respect to time and applying
of the Pauli-Schrodinger equation with Hamiltonian (\ref{H}) leads
to continuity equation   in the rotating frame

\begin{equation} \label{n1}
\partial_t n(\mathbf{r},t) +\nabla \mathbf{j}(\mathbf{r},t)=0, \end{equation}
where the current density vector takes a form of

\begin{equation} \label{J}
\mathbf{j}(\mathbf{r},t)=\Biggl\langle\frac{1}{2m_{p}}
\Biggl((\mathbf{\hat{j}}_{p}\Psi)^{+}\Psi +\Psi^{+}\mathbf{\hat{j}}_{p}\Psi\Biggr)\Biggr\rangle. \end{equation}

In the definition for the   current density vector (\ref{J}) we use the current operator of i-th particle in the form

     \begin{equation} \label{J1}
\mathbf{\hat{j}}_p=\mathbf{\hat{D}}_p-m_p\mathbf{\Omega}\times\mathbf{r}_p-\frac{\mu_p}{c}\hat{\mathbf{\sigma}}_p\times\end{equation}

$$ \times
\Biggl(\mathbf{E}_{p,ext}+\frac{(\mathbf{\Omega}\times\mathbf{r}_p)}{c}\times\mathbf{B}_{p}\Biggr)+\frac{1}{2} \sum^{N}_{p\neq n} \frac{q_p}{c}\mu_p\varepsilon^{\alpha\beta\gamma}\hat{\sigma}_p^{\beta}\mathbf{\nabla}_p^{\gamma} G_{pn}.
$$

We derive the velocity of i-th particle $\mathbf{v}_p$ is determined by
equation

         \begin{widetext}
       \begin{equation}\label{v}   \mathbf{v}_p=\frac{1}{m_p}\Biggl(\mathbf{\nabla}_pS-\frac{q_p}{ c}\mathbf{A}_p\Biggr)-\mathbf{\Omega}\times\mathbf{r}_p
        -\frac{\mu_p}{cm}\hat{\mathbf{\sigma}}_p\times
\mathbf{E}_{p,eff}-\frac{i\hbar}{m_p}\phi^{+}\nabla_p\phi, \end{equation}\end{widetext} where the second term in the definition (\ref{v}) is the rotational velocity characterizing
           the effect of mechanical
           rotation. The effective electric field $\mathbf{E}_{eff}$ consists of the external electric field, the Coulomb
           internal field and the internal electric field originating from the mechanical rotation. It was shown that this field effects
           due to the spin-orbit interaction SOI with the mechanical rotation and can be displayed in the large SOI systems.   In general case
$\mathbf{v}_p(R,t)$ depends on the coordinate of all particles of the system
$R$, where $R$ is the totality of $3N$ coordinates of $N$ particles
of the system $R=(\mathbf{r}_{1},...,\mathbf{r}_{N})$.  The $S(R, t)$ value in the formula (\ref{v}) represents the phase of the wave function

          \begin{equation}\label{psi}
\Psi(R,t)=a(R,t)\exp\Biggl(\frac{iS(R,t)}{\hbar}\Biggr)\phi(R,t),
\end{equation} where $\phi$, normalized such that $\phi^+\phi=1$, is the new spinor.
Velocity field $\mathbf{v}(\mathbf{r},t)$ is the velocity of the local centre of
mass in the rotating frame, is measured in the inertial frame and
 indicates a quantity measured in the rotating frame

    \begin{equation}\label{v1}
\mathbf{j}(\mathbf{r},t)=n(\mathbf{r},t)\mathbf{v}(\mathbf{r},t),
\end{equation}  where the quantum equivalent of the thermal speed is determined as
$\mathbf{u}_p(\mathbf{r},R,t)=\mathbf{v}_p(R,t)-\mathbf{v}(\mathbf{r},t).$ We differentiate the momentum density (\ref{J})
with respect to time and apply the many-particle
Pauli-Schrodinger equation  to time derivatives of the wave functions $\Psi(R,t)$. As a result, a momentum balance equation can
be obtained in the form

\begin{equation} \label{J2}
\partial_t\mathbf{j}(\mathbf{r},t)+\frac{1}{m}\partial_{\beta}\mathbf{\Pi}^{\beta}(\mathbf{r},t)=\mathbf{F},
\end{equation}      where
\begin{equation}  \label{P}
\mathbf{\Pi}^{\beta}(\mathbf{r},t)=\Biggl\langle\frac{1}{4m_{p}}
\Biggl((\mathbf{\hat{j}}_{p}\Psi)^{+}\hat{j}_p^{\beta}\Psi +\Psi^{+}(\mathbf{\hat{j}}_{p}\hat{j}^{\beta}_p\Psi)+c.c.\Biggr)\Biggr\rangle\end{equation}
represents the momentum current density tensor and $\mathbf{F}$ represents a force field including the Coriolis, centrifugal and Euler forces
in the rotating frame. We introduce the separation of particles' thermal movement with
velocities $\mathbf{u}_p(\mathbf{r},R,t)$ and the collective
movement of particles with velocity $\mathbf{v}(\mathbf{r}, t)$ in equations of
continuity (\ref{n1}) and of the momentum balance (\ref{J2}).  For that we substituted the
wave function (\ref{psi}) in the definition of the basic hydrodynamical quantities. The
momentum current density tensor (\ref{P}) will has the new form of

$$
\mathbf{\Pi}^{\beta}(\mathbf{r},t)=mn(\mathbf{r},t)\textbf{v}(\mathbf{r},t)v^{\beta}
(\mathbf{r},t)$$
            \begin{equation} \qquad\qquad\qquad\qquad +\mathbf{p}^{\beta}_{thermal}(\mathbf{r},t)+\mathbf{\Upsilon}_{quantum}^{\beta}(\mathbf{r},t).\end{equation}

As we can see, the kinetic pressure tensor and quantum pressure tensor appear in the definition for the momentum current density tensor.

\begin{equation}
\mathbf{p}^{\beta}_{thermal}(\mathbf{r},t)=\Biggl\langle a^{2}
m_p\cdot\textbf{u}_p u_p^{\beta}\Biggr\rangle \end{equation}                  is the tensor of kinetic pressure and

\begin{equation}  \label{Bohm}
\mathbf{\Upsilon}_{quantum}^{\beta}(\mathbf{r},t)=\Biggl\langle
-a^{2}\frac{\hbar^{2}}{2m_p}\frac{\partial^{2}\ln a}{\partial\textbf{r}_p \partial r_p^{\beta}}+\frac{\hbar^2a^2}{4m_p}
\nabla \mathbf{s}_{p}\cdot\nabla^{\beta}\mathbf{s}_p\Biggr\rangle, \end{equation}
   where the first term is the Madelung  quantum potential. This tensor  is proportional to $\hbar^{2}$,
   has a purely quantum origin and can therefore be interpreted as an additional quantum pressure. The second term
   characterizes the force produced by the self-interactions of the spins $\mathbf{s}_p$.

   Taken in the approximation
of self-consistent field, the  continuity
equation and momentum balance equation in the rotating frame have a form
     \begin{widetext}
\begin{equation} \label{n3}
\partial_{t}n(\mathbf{r},t)+\mathbf{\nabla}(n\textbf{v})(\mathbf{r},t)=0,
\end{equation}

$$
mn(\mathbf{r},t)(\partial_{t}+\mathbf{v}\mathbf{\nabla})\mathbf{v}(\mathbf{r},t)=
en(\mathbf{r},t)\mathbf{E}(\mathbf{r},t)+\frac{e}{c}n(\mathbf{r},t)\mathbf{v}(\mathbf{r},t)\times\mathbf{B}(\mathbf{r},t)
-
\nabla_{\beta}\mathbf{p}^{\beta}(\mathbf{r},t)$$

$$-\frac{\hbar^2}{2m}n(\mathbf{r},t)\mathbf{\nabla}\Biggl(\frac{\triangle\sqrt{n(\mathbf{r},t)}}{\sqrt{n(\mathbf{r},t)}}
\Biggr)+\mu n(\mathbf{r},t)s_{\beta}(\mathbf{r},t)\mathbf{\nabla}B^{\beta}(\mathbf{r},t)-\frac{\hbar^2}{4m}\partial_{\beta}\Biggl(n\nabla\mathbf{s}(\mathbf{r},t)
\cdot\partial^{\beta}\mathbf{s}(\mathbf{r},t)\Biggr)$$

 \begin{equation}\label{j3} \qquad\qquad\qquad\qquad\qquad\qquad\qquad\qquad\qquad +\frac{\hbar}{2}n(\mathbf{r},t)s_{\beta}(\mathbf{r},t)\mathbf{\nabla}\Omega^{\beta}(\mathbf{r},t)+\mathbf{F}_{inertial}(\mathbf{r},t)+
\mathbf{F}_{SO}(\mathbf{r},t),\end{equation}\end{widetext} where $n$, $m$ and $\mathbf{v}$ denote the density, mass and fluid velocity, $\mathbf{E}$ and $\mathbf{B}$
are the  electric and magnetic fields in the rotating frame, $\mathbf{p}^{\beta}$ is the thermal pressure tensor,
$\mathbf{s}^2=1-<a^2\mathbf{\xi}_p\cdot\mathbf{\xi}_p>/n$ is the macroscopic spin angular momentum, where $\mathbf{\xi}_p$ is the
thermal fluctuations of the spin about the macroscopic average.  The first two terms describe the interaction
with external electromagnetic field. As we can see, the transformations to the rotating frame of
reference does not change the continuity equation (\ref{n3}).  The first term at the
right side of the equation  (\ref{j3})
represents the effect of external electric field on the charge
density and the second term represents the Lorentz
force in the rotating frame.  The fourth term at the
right side of the equation (\ref{j3}) is a quantum force produced by quantum Madelung potential.  The fifth term is the effect of non-uniform
magnetic field on the
magnetic moment.    The fifth term appears in the equation of motion (\ref{j3}) through the
magnetization energy and represents the Stern-Gerlach force due to the coupling between magnetic moment and magnetic field. The sixth term at the
right side of the equation (\ref{j3}) is the  spin
force, produced by the spin stress \cite{230}. The spin stress in the one-particle model
was derived by Takabayasi \cite{23}. This development was performed using the  postulate  that a corpuscle of mass is embedded in
the spinor wave.
Spinor, in that way,  is represented in terms of the  spacetime-dependent
Euler angles  which define the orientation of the triad relative  to the fixed set of Cartesian axes \cite{24}, \cite{25}.
In the context of this representation the {\em spinor wave must constitute  a new form of physical field that affects on the
corpuscle of mass moving within it}.

The seventh  term on the right side of equation (\ref{j3}) represents the spin-rotation coupling.
The eighth term is the  inertial force density in the rotating frame

     $$
     \mathbf{F}_{inertial}(\mathbf{r},t)=-2m\mathbf{\Omega}\times n(\mathbf{r},t)\mathbf{v}(\mathbf{r},t)-
     m\frac{\partial\mathbf{\Omega}}{\partial t}\times \mathbf{P}(\mathbf{r},t)
                              $$

           \begin{equation} \label{inertia}   \qquad\qquad- m\mathbf{\Omega}\times(\mathbf{\Omega}\times\mathbf{P}(\mathbf{r},t))
              -m\nabla_{\beta}\mathbf{\Omega}\times\mathbf{\Lambda}^{\beta}(\mathbf{r},t),  \end{equation}

              where the displacement vector
                                                                              \begin{equation}  \label{disp}
                \mathbf{P}(\mathbf{r},t)=\Biggl\langle \Psi^+\mathbf{r}_p\Psi\Biggr\rangle
                \end{equation}  and
                $$   \mathbf{\Lambda}^{\beta}(\mathbf{r},t)=
                \Biggl\langle\frac{1}{2m_{p}}\mathbf{r}_p
\Biggl((\hat{j}^{\beta}_{p}\Psi)^{+}\Psi +\Psi^{+}\hat{j}^{\beta}_{p}\Psi\Biggr)\Biggr\rangle.
                $$

The force field (\ref{inertia}) depends on the mechanical rotation  velocity $\mathbf{\Omega}$ and leads from the coupling between the mechanical rotation and angular momentum (\ref{H2}). The first term at the
right side of the (\ref{inertia}) is the {\em Coriolis} force density, the third term represents the {\em centrifugal}
force density, the second and forth terms form the {\em Euler} force field density.   The main features of the Coriolis force
that this force
cannot do work on the fluid and strives to deflect a fluid particle in a direction
perpendicular to its instantaneous velocity.    The last term on the right side of equation (\ref{j3}) is the spin-orbit force field density.

\subsection{Equation for the evolution of displacement vector}

To close the QHD equations set (\ref{n3}), (\ref{j3}) we derive
equation for the displacement evolution. If we differentiate
the definition for displacement (\ref{disp}) with respect to time
and apply the Schr¨odinger equation, the required equation
for the displacement evolution can be obtained

                       \begin{equation} \label{Pol}
\partial_t \textbf{P}(\textbf{r},t)+\partial_{\beta} \mathbf{\Lambda}^{\beta}(\textbf{r},t)=0,
\end{equation}

We have two ways to close the QHD equations set. The
first one is to express $\mathbf{\Lambda}^{\beta}$ in terms of $n$, $\upsilon$
and $\textbf{P}$ using additional assumptions or experimental
data. The other way is to derive the equation for evolution
$\mathbf{\Lambda}^{\beta}$ in the same fashion it was accomplished
previously for other material fields. Now the
evolution equation $\mathbf{\Lambda}^{\beta}$ occurs in the form of

\begin{widetext}
\begin{equation}    \label{Pol2} \partial_{t}\mathbf{\Lambda}^{\beta}(\textbf{r},t)+\frac{1}{m}\partial_{\gamma}\mathbf{\Lambda}^{\beta\gamma}(\textbf{r},t)= \frac{e}{m}\textbf{P}(\textbf{r},t)E^{\beta}_{ext}(\textbf{r},t)
+\frac{e}{mc}\epsilon^{\beta\gamma\delta}\mathbf{\Lambda}_{\beta}(\textbf{r},t)B^{\delta}(\textbf{r},t)-
 2\mathbf{\Omega}\times\mathbf{\Lambda}(\textbf{r},t)$$

$$\qquad\qquad\qquad\qquad\qquad\qquad\qquad-\mathbf{\Omega}\times(\mathbf{\Omega}\times\textbf{P}(\textbf{r},t)\cdot \frac{P^{\alpha}(\textbf{r},t)}{n(\textbf{r},t)})-\partial_t\mathbf{\Omega}\times\textbf{P}(\textbf{r},t)\frac{P^{\alpha}(\textbf{r},t)}{n(\textbf{r},t)}\end{equation}
\end{widetext}

The first two terms describe
the interaction of particles with external electromagnetic
field. The next terms represent the inertial forces.

\subsection{Equation for the spin evolution}

  To close the  equations set (\ref{n3}) and  (\ref{j3}) we derive
equation for the spin   evolution. If we differentiate
the definition for spin-polarization

                  \begin{equation}
\mathbf{s}(\mathbf{r},t)=\Biggl\langle\phi^{+}\hat{\mathbf{\sigma}}_p\phi\Biggr\rangle
\end{equation}
 with respect to time
and apply the Pauli-Schrodinger equation with the Hamiltonian in the rotating frame, the required equation
for the spin  evolution $\mathbf{s}$ without thermal effects can be obtained

 \begin{widetext} $$
(\partial_{t}+\mathbf{v}\mathbf{\nabla})\mathbf{s}(\mathbf{r},t)=\frac{2\mu}{\hbar}\mathbf{s}(\mathbf{r},t)\times\mathbf{B}(\mathbf{r},t)
+\mathbf{s}(\mathbf{r},t)\times\mathbf{\Omega}(\mathbf{r},t)
          $$
            \begin{equation}  \label{M3}\qquad\qquad\qquad+\frac{\hbar}{2mn(\mathbf{r},t)}\mathbf{s}(\mathbf{r},t)\times\partial_{\beta}(n(\mathbf{r},t)\cdot\partial^{\beta}\mathbf{s}(\mathbf{r},t))
            -\frac{2\mu}{\hbar cn(\mathbf{r},t)}\varepsilon^{\alpha\beta\gamma}\varepsilon_{\beta\mu\nu}E^{\mu}_{eff}(\mathbf{r},t)J^{\gamma\nu}_M(\mathbf{r},t)
                 \end{equation}    \end{widetext}

The first term at the
right side of the equation (\ref{M3})  represents the torque caused by the interaction with the external magnetic field
and the magnetic field of the spin-spin interparticle interactions in the rotating frame. The second term is the spin rotation coupling term in the rotating frame and can
be interpreted as the torque caused by the interaction with effective magnetic field $\mathbf{B}_{\Omega}$ introducing the Barnett effect to the right side of equation (\ref{M3}).
The third term is the spin torque \cite{23}.  The fourth term characterizes the torque resulting from   spin-orbit coupling in the effective electric field. The magnetic moment flux tensor occurs in this equation
(\ref{M3}) in the form

  \begin{equation}  \label{JM} \mathbf{J}^{\beta}_{M}(\mathbf{r},t)=\Biggl\langle\frac{1}{4m_{p}}\Biggl(\Psi^{+}\hat{\mathbf{\sigma}}_p\hat{j}^{\beta}_{p}\Psi+
(\hat{\mathbf{\sigma}}_p\hat{j}^{\beta}_{p}\Psi)^{+}\Psi\Biggr)\Biggr\rangle. \end{equation}

We introduce the thermal fluctuation $\mathbf{w}_p(\mathbf{r},R,t)=\mathbf{s}_p(R,t)-\mathbf{s}(\mathbf{r},t)$
of the spin about the macroscopic average $\mathbf{s}(\mathbf{r},t)$
which determined  in the neighborhood
of $\mathbf{r}$ in a physical space. Using the Madelung decomposition of the N-particle wave function (\ref{psi})
we have obtained the definition for the magnetic moment flux tensor
in the rotating frame

   \begin{equation}  \label{JM1}
\mathbf{J}^{\beta}_{M}(\mathbf{r},t)=\mathbf{s}(\mathbf{r},t)n(\mathbf{r},t)v^{\beta}+\mathbf{j}^{\beta}_{thermal}(\mathbf{r},t) \end{equation}       $$\qquad\qquad\qquad - \frac{\hbar}{2m}n(\mathbf{r},t)\mathbf{s}(\mathbf{r},t)\times\partial_{\beta}\mathbf{s}(\mathbf{r},t)
$$
where we introduce the definition for the thermal flux density tensor

    \begin{equation}
\mathbf{j}^{\beta}_{thermal}(\mathbf{r},t)=\Biggl\langle a^{2}
\cdot\textbf{s}_p u_p^{\beta}\Biggr\rangle. \end{equation}

We have two ways to close the  equations set. The
first one is to express $\mathbf{J}^{\beta}_{M}(\mathbf{r},t)$ in terms of $n(\mathbf{r},t)$,
$\mathbf{v}(\mathbf{r},t)$ and $\mathbf{s}(\mathbf{r},t)$ using additional assumptions or experimental
data. The other way is to derive the equation for evolution
$\mathbf{J}^{\beta}_{M}$ in the same fashion it was accomplished
previously for other material fields.

\subsection{The evolution of polarization in the rotating frame}

The evolution of polarization for the particles with the electric dipole moment (EDM) can be derived using a method of quantum hydrodynamics. We receive the equations for the system
of charged particles with EDM. We research the fluid of EDM-having polar particles (polar molecules).  Obtaining equations
could be used for neutral particles with the EDM as well.
Method of quantum hydrodynamics allows to present dynamic of system
of interacting quantum particles in terms of functions defined
in 3D physical space in the rotating reference frame.  The  many-particle  Hamiltonian for the EDM-having particles in the rotating reference frame  has the form

\begin{equation}\label{HH} \hat{H}=\hat{H}_{0}+\hat{H}_{rotor}\end{equation}

  $$  \hat{H}_0=\sum^{N}_{p=1}\Biggl({\frac{\hat{\mathbf{D}}^{2}_{p}}{2m_p}+q_p\varphi_{p,ext}-\mathbf{d}_{p}\mathbf{E}_{p,ext}}\Biggr)
$$
\begin{equation} \label{H4}\qquad\qquad\qquad+\frac{1}{2}\sum^{N}_{p\neq n} q_pq_nG_{pn}-\frac{1}{2}\sum^{N}_{p\neq n,n}G^{\alpha\beta}_{pn}\hat{d}^{\alpha}_{p}\hat{d}^{\beta}_{n},   \end{equation}

 \begin{equation} \label{H5}\hat{H}_{rotor}= -\sum^{N}_{p=1}\Biggl(\varepsilon^{\alpha\beta\gamma}\Omega^{\alpha}_p\cdot r^{\beta}_p\hat{D}^{\gamma}_p +\frac{\hbar}{2}\hat{\sigma}^{\alpha}_{p}\cdot\Omega^{\alpha}_{p}\Biggr). \end{equation}

 The new third term in (\ref{H4})  is considered in the Hamiltonian  of
particles through the dipole energy in the external electric field.   The fourth term in (\ref{H4}) presents the Coulomb interaction between charged particles and the last term  characterizes the dipole-dipole interactions between dipoles, where $G^{\alpha\beta}_{pn}=\partial^{\alpha}_{p}\partial^{\beta}_{n}/r_{pn}.$

From the many-particle  Hamiltonian (\ref{HH}) the required equation
for the polarization evolution can be obtained as \cite{192}

\begin{align} \label{P}
\partial_t \textbf{D}(\textbf{r},t)+\partial_{\beta} \textbf{R}^{\beta}(\textbf{r},t)=0,
\end{align}
where the polarization vector field of the EDM-having particles has the form
\begin{equation} \label{P1}
\textbf{D}(\mathbf{r},t)=\langle\Psi^{+}\textbf{d}_p\Psi\rangle,
\end{equation}
and a polarization current can be derived in the form

\begin{align}
\textbf{R}^{\beta}(\textbf{r},t)=\Biggl\langle\frac{\textbf{d}_p}{2m_{p}}({\hat{j}^{+\beta}_{p}\Psi^{+}\Psi+\Psi^{+}\hat{j}^{\beta}_{p}\Psi})\Biggr\rangle.
\end{align}

         \section{Wave of polarization}

\subsection{The inertial frame}

We investigate the system of neutral particles resides in
a uniform electromagnetic field. It is also assumed that
interactions make the largest contribution into the changes
in $\textbf{R}^{\beta}$ \cite{192}.
If so then we use the equation (\ref{P}) and the equation for the polarization current density evolution \cite{192}

        $$
\partial_t \textbf{R}^{\beta}(\textbf{r},t)=\sigma\frac{\textbf{D}(\textbf{r},t)D^{\gamma}(\textbf{r},t)}{mn(\textbf{r},t)}\times
     $$ \begin{equation}\label{RR2}    \qquad\qquad
\times\nabla_{\beta}\int d\textbf{r}^{'}G^{\gamma\mu}(\textbf{r},\textbf{r}^{'})D_{\mu}(\textbf{r}^{'},t)\end{equation}

\begin{figure}
\centering
\subfigure[] {
\includegraphics[ width=75mm]{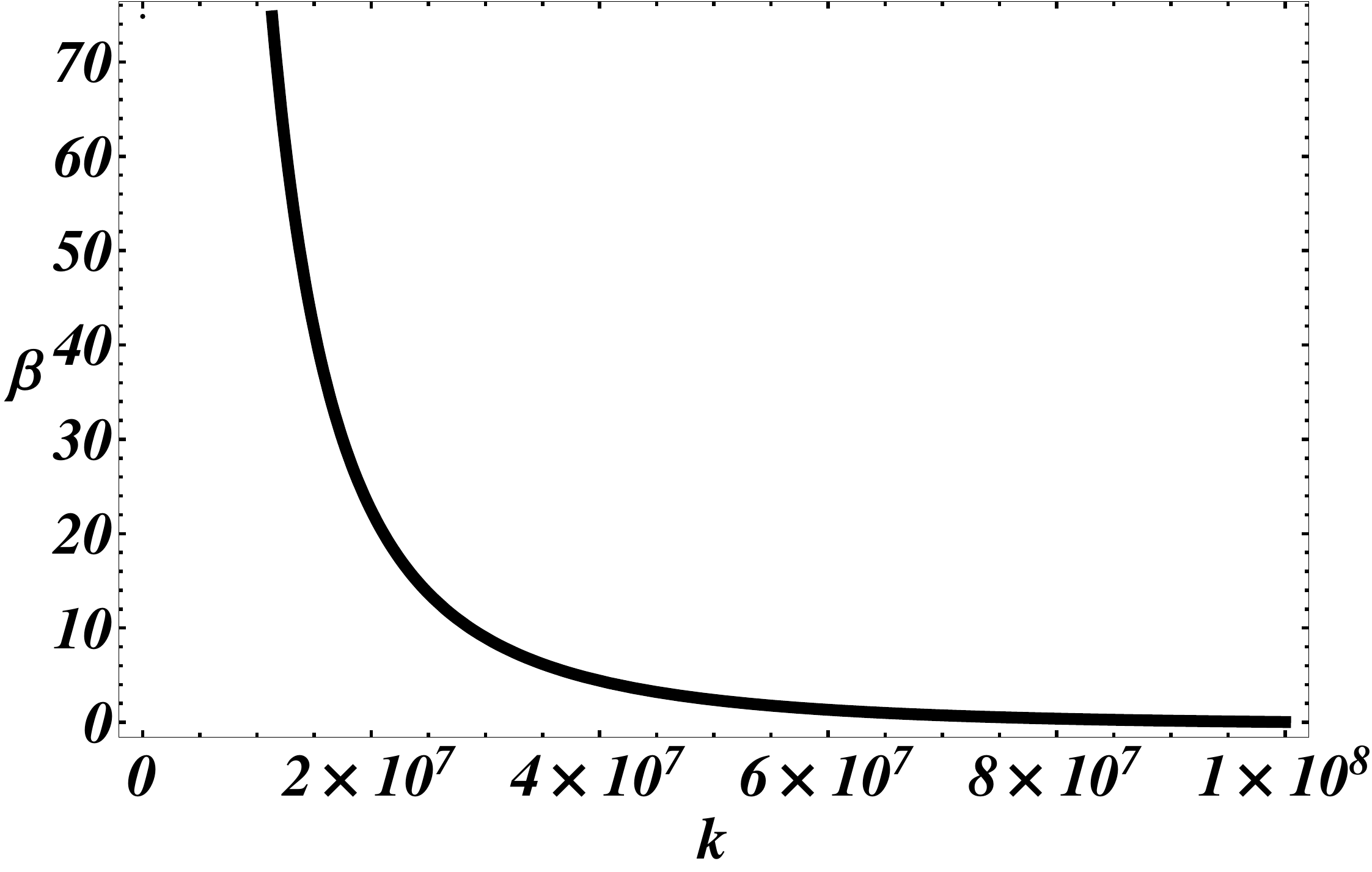}\label{subfig1}}
\subfigure[]{
\includegraphics[width=75mm]{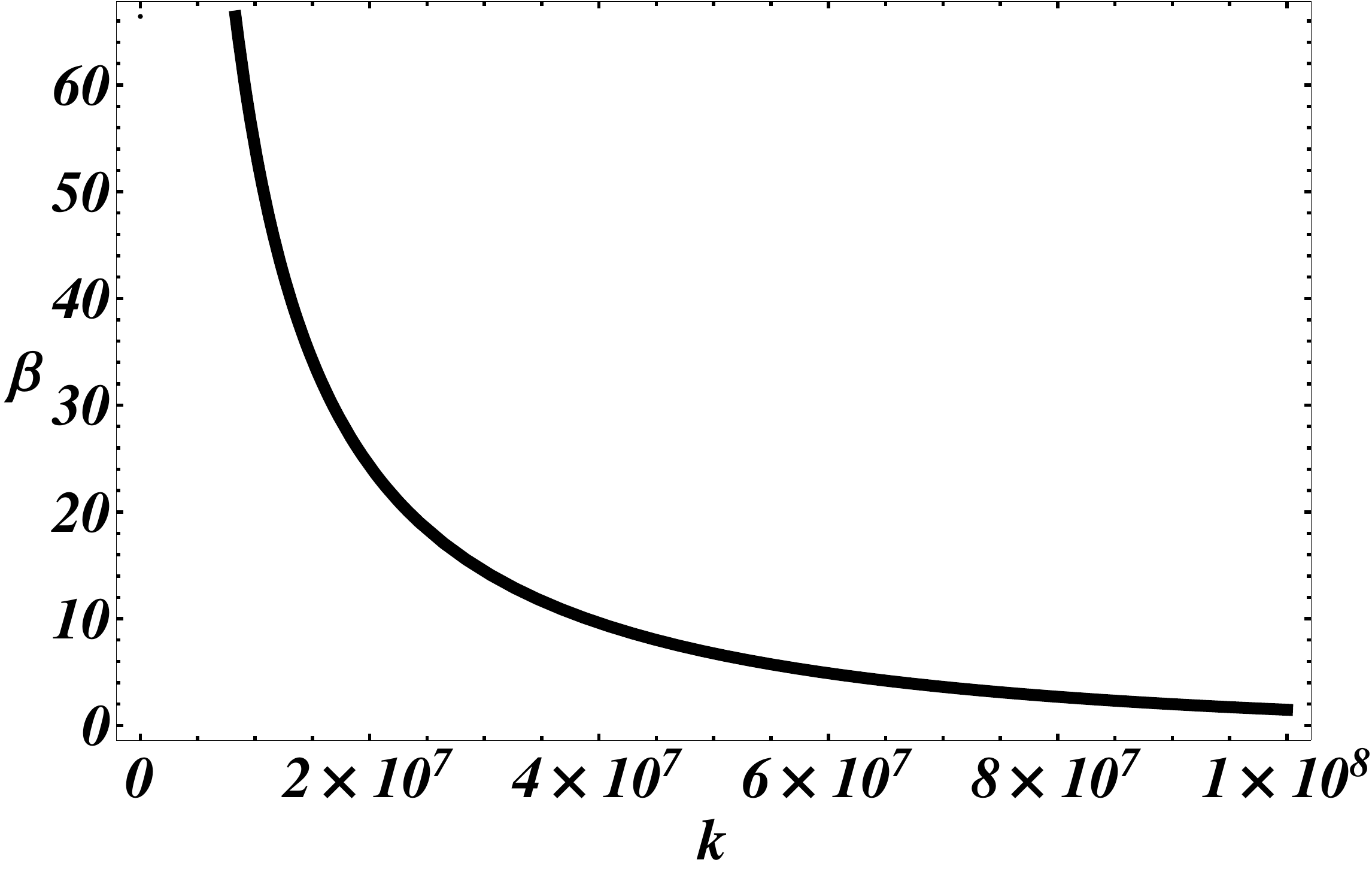}\label{b}}
\caption{Color online:  The figure (a) presents of the variable $\beta_{1}(k)$ on the wave
vector k. Ionic or molecular radius $r_0$ are assumed to equal $0.1$ $nm$. The figure (b) presents the dependence of the variable $\beta(k)$
on the wave vector k. Ionic or molecular radius $r_0$ is assumed to
equal 0.1 nm.}
\end{figure}

The term on the right hand side of (\ref{RR2}) characterizes the dipole-dipole interactions between the particles. This
allows the analysis of polarization waves in a system of
neutral particles.  If we derive a solution for eigenwaves in a
2D system the dispersion equation has a form of

\begin{equation}\label{w}\omega=\sqrt{\sigma\frac{\beta(k) }{mn_{0}}}\mid\kappa\mid E_{0}k^{3/2},\end{equation}  where $\beta(k)$ is defined by the relation
\begin{equation}\label{di const two dim}\beta(k)=2\pi\int_{\xi}^{\infty}dr\frac{J_{0}(r)}{r^{2}},\end{equation}
here $\xi=r_{0}k$, $r_{0}$ there is an ionic or molecular radius and $k=\sqrt{k_{x}^{2}+k_{y}^{2}}$ is a modulus of the wave vector. As $\lambda_{min}=2\pi/k_{max}>2r_{0}$ then $\xi\subset(0,\pi)$.
    The dispersion dependence (\ref{w}) is presented on fig. (\ref{subfig1}).

In 1D case $\omega(k)$ occurs as
\begin{equation}\label{di disp for neutral part one dim}\omega=\sqrt{\frac{\sigma\beta_{1}(k)}{mn_{0}}}\mid\kappa\mid E_{0}k^{2},\end{equation}
where
\begin{equation}\label{di koef one dim}\beta_{1}(k)=2\int_{\xi}^{\infty}dr\frac{cos(r)}{r^{3}}.\end{equation}
The quantity (\ref{di koef one dim})   is presented on Figs. (\ref{subfig1}).  The dispersion relations (\ref{w}) and  (\ref{di disp for neutral part one dim}) characterize the waves of polarization in the system of neutral particles with dipole moments. This waves exist on a level with the acoustic waves. Equations of continuity (\ref{n3}) and of the momentum balance (\ref{j3}) herein describe the dynamics of the acoustic wave.  Dispersion branches of a novel type that occurs due to the
polarization dynamics were discovered in various  physical
systems \cite{192}. The waves of  electric polarization we
discovered possess the following feature -  their frequencies
$\omega$ tends to zero provided that $k \rightarrow 0$.
    \begin{figure}
\centering
\subfigure[] {
\includegraphics[ width=72mm]{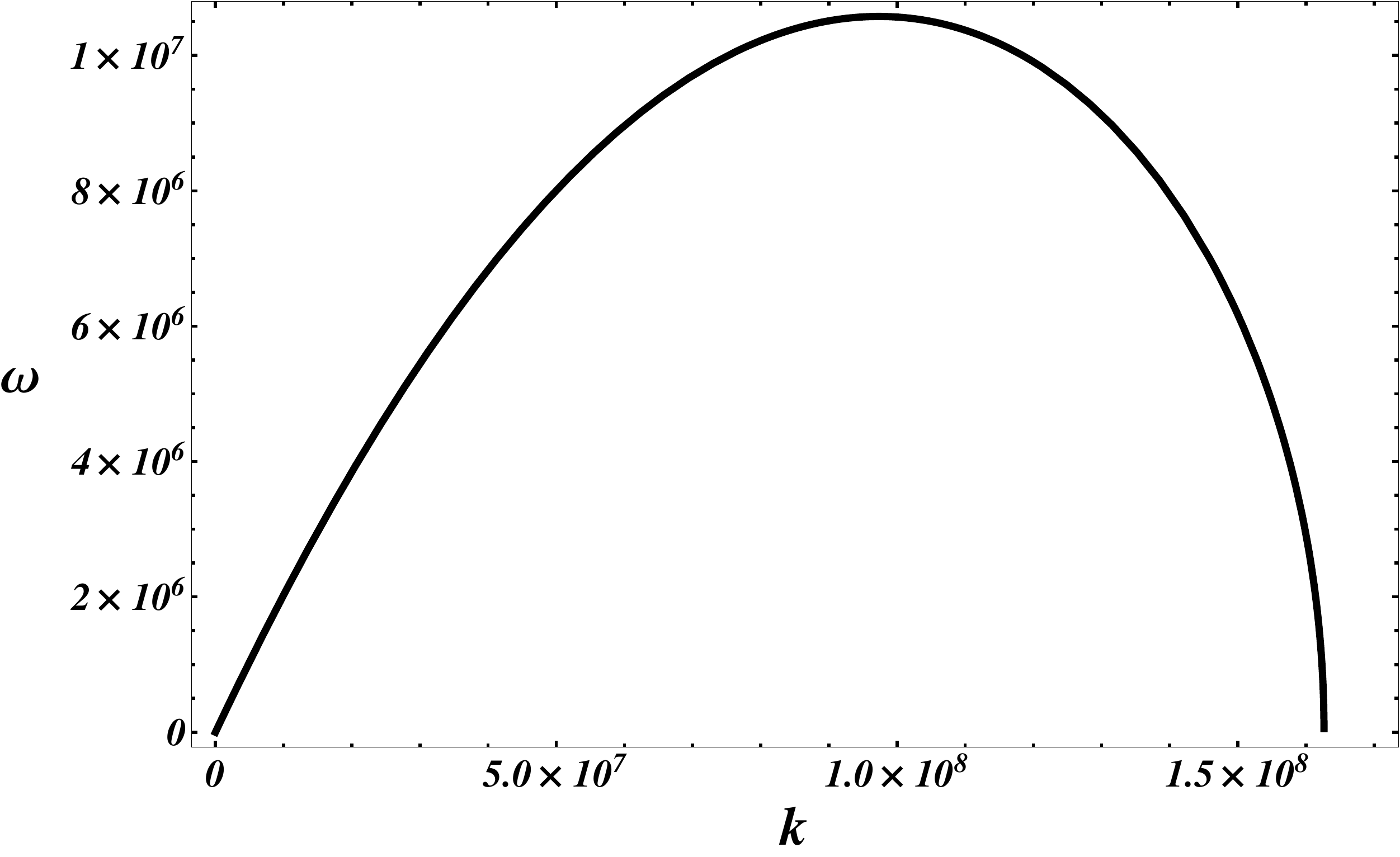}\label{subw1}}
\subfigure[]{
\includegraphics[width=75mm]{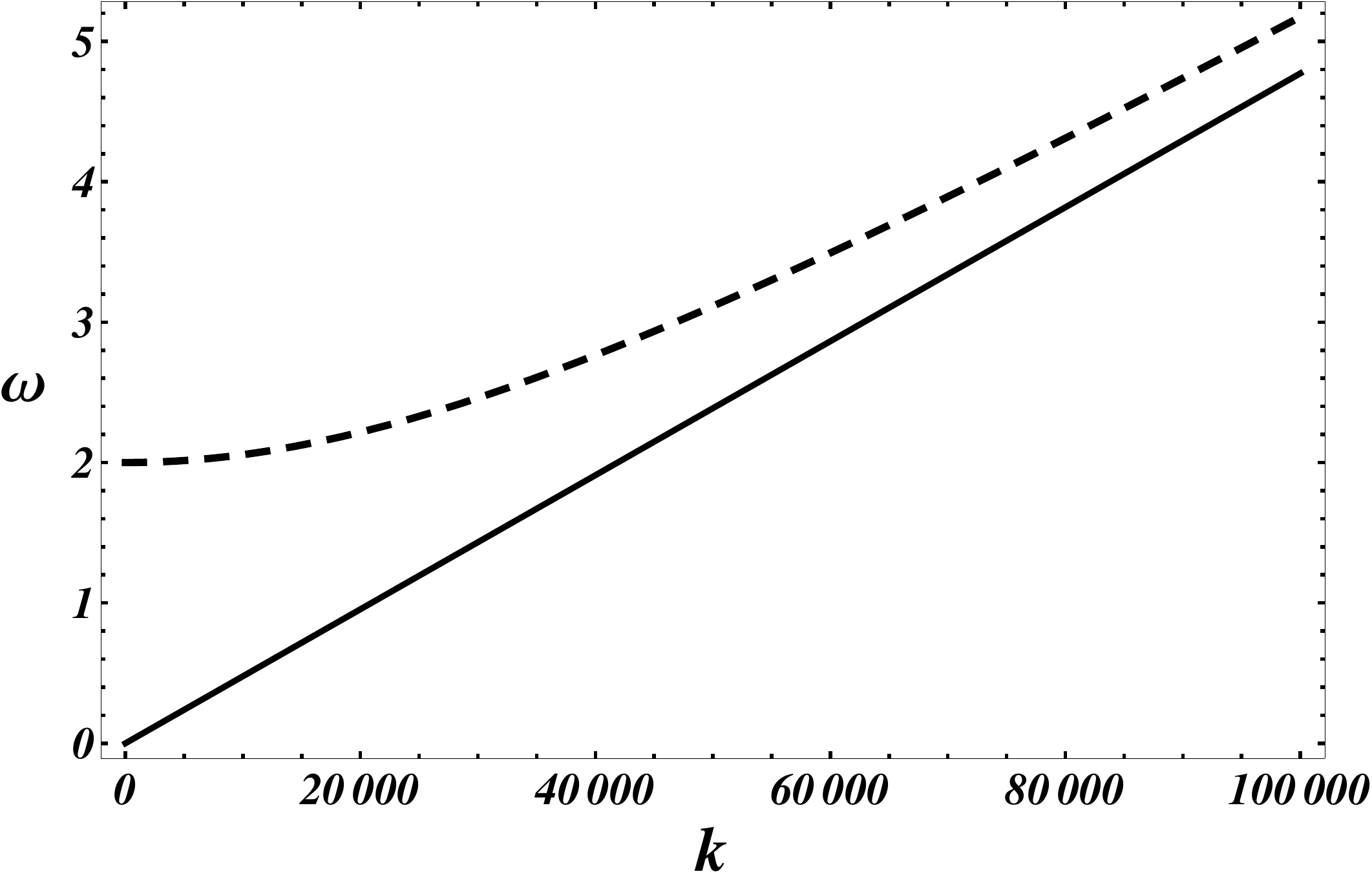}\label{subw2}}
\caption{Color online:   The dependence of frequency $\omega$ on the wave vector $k$
is displayed for the case of two-dimension polarization mode
which dispersion characteristic is defined by the equation (\ref{w}) for the figure (a) and of two-dimension polarization mode modified by rotation for the figure (b). The dashed branch describes the polarization wave modified by rotation or dipole-inertial 2D-mode (\ref{w4}).
The  radius $r_0$ is supposed to be 0.1 nm. Equilibrium
polarization has form $D_0 = \kappa E_0$. Static electric permeability
$\kappa$ is defined by the equation $\kappa = n_0d^2_0
/(3k_BT)$, where $d_0$ - is
a dipole moment of an molecule, $T$ - temperature of the medium,
$k_B$ - Boltzmann constant. System parameters are assumed to be
as follows: $n_0$ = 10$^{12}$sm$^-2$, $d_0\simeq 0.16 D$, $T = 100K$,
$E_0\simeq10^{2}V/m$ and $m = 10^{-23}g$.}
\end{figure}

\subsection{Dipole-inertial wave}
   \subsubsection{2D-eigenwaves}
In this section we consider the fluid
of neutral particles in the rotating reference frame.
  We derive dispersion
characteristics of eigenwaves in 2D systems of neutral
 particles that take account of the EDM collective
dynamics. The
particles are assumed here to localize on a $xy$ -
plane and reside in
a uniform electromagnetic field orthogonal to the plane $xy$.   For simplicity the centrifugal force is neglected and only
the effects of the Coriolis force are taken into consideration $|\mathbf{\Omega}\times(\mathbf{\Omega}\times\mathbf{P})| << 2|n\mathbf{\Omega}\times \mathbf{v}|$ the equation (\ref{RR2}) transforms into

              $$
\partial_t \textbf{R}^{\alpha}(\textbf{r},t)=\sigma\frac{D^{\alpha}(\textbf{r},t)D^{\gamma}(\textbf{r},t)}{mn(\textbf{r},t)}\times
     $$ \begin{equation}\label{RR4}
\times\nabla\int d\textbf{r}^{'}G^{\mu\gamma}(\textbf{r},\textbf{r}^{'})D_{\mu}(\textbf{r}^{'},t)-2\mathbf{\Omega}\times\textbf{R}^{\alpha},\end{equation}
      and to close the  equation  (\ref{RR4}) we use the
equation for the polarization evolution (\ref{P}). If we derive a solution for eigenwaves in a
2D system the dispersion equation has a form of

\begin{equation}\label{w4}\omega=\sqrt{\sigma\frac{\beta(k) }{mn_{0}}\mid\kappa\mid^2 E_{0}^2k^{3}+4\Omega^2},\end{equation}
    the dispersion relation  characterizes the 2D-polarization wave modified by rotation (fig. \ref{subw2}), where the $z$ axis is directed along the rotation axis $\mathbf{\Omega}=\Omega\textbf{z}.$

\subsubsection{3D-wave}
To
analyze 3D systems we use the set of equations (\ref{P}) and the polarization current density equation. The equation for the
polarization current density in the rotating frame and in the approximation of the self-consistent field has the form of

       \begin{equation}\label{RR3}
\partial_t                     \textbf{R}^{\alpha}=\sigma\frac{D^{\alpha}D^{\gamma}}{mn}
\times\nabla E^{\gamma}-2 \mathbf{\Omega}\times\textbf{R}^{\alpha},\end{equation}
           where the total electric field $\textbf{E}$ leads from  a field equation   $ \nabla\cdot\textbf{E}=-4\pi \nabla\cdot\textbf{D}.$  Here $z$ axis is directed along the rotation axis, $\textbf{D}_0=D_{\parallel}\textbf{z}$ and two-dimensional position vector
   $\mathbf{r}_{\perp}$ is directed in the $xy$ plane normal to the rotation axis, a wave number $k_{\parallel}$ is the component of the
   three-dimension wave vector along the rotation axis and $\mathbf{k}_{\perp}$ is the wave vector in the $xy$ plane. These equations give the dispersion low

                  \begin{equation}\label{w1}\omega^2= \frac{\Omega^2_d+4\Omega^2}{2}\pm\end{equation}
                  $$\pm\sqrt{\biggl(\frac{\Omega^2_d}{2}+2\Omega^2\biggr)^2-4\Omega^2_{d}\Omega^2\cos^2\theta}.
                  $$

                   For a fluid at rest ($\Omega=0$) this relation gives the frequencies the zero-frequency transverse mode of a fluid at rest $\omega=0$ and $\omega=\Omega_d$ corresponding to the transverse polarization wave.   It seems interesting to analyze the effect of rotation on the wave propagation in  different limits. Obviously, in the approximation of large wave vectors $k_{\parallel}\rightarrow \infty$ and $\Omega^2_d>>\Omega^2$ one branch of dispersion yields the inertial wave 
                       \begin{equation}\omega^2=4\Omega^2\frac{k^2_{\parallel}}{k^2},\end{equation}
and another represents the polarization wave modified by rotation

                                  \begin{equation}\label{w2}\omega^2=\sigma\frac{4\pi D^2_0}{mn}k^2_{\parallel}-4\Omega^2\frac{k^2_{\parallel}}{k^2}\rightarrow \Omega^2_d.\end{equation}



\subsection{Conclusions}

In this paper  the method of QHD was developed for spin-bearing and EDM-bearing particles in the rotating frame.
QHD equations are a consequence of MPSE in which particles'
interaction is directly taken into account.  The system of QHD equations we constructed
is comprised by equations of continuity (\ref{n3}), of the momentum
balance (\ref{j3}) and  of the spin evolution equation (\ref{M3}). Using the developed method we derived the inertial force field (\ref{inertia}) which consists of the {\em Coriolis} force density,  the {\em centrifugal}
force density and of  the {\em Euler} force field density. We close our system of equations by the displacement evolution equation (\ref{Pol2}).    In our studies of wave processes we used a
self-consistent field approximation of the QHD equations. In this paper we also  analyzed wave excitations caused by
EDM dynamics in systems of charged and neutral particles in the rotating frame.
For this purpose we derived the polarization evolution equation (\ref{P}) and  the polarization
current  for 2D systems (\ref{RR4}) and 3D fluids (\ref{RR3}).
Waves in a 2D fluid of EDM-bearing neutral particles  in various physical
dimensions were considered.  The effect of mechanical rotation on the  polarization dynamics was derived.  Dispersion branches of a novel type that occurs due to
polarization dynamics in the rotating frame were discovered for 2D (\ref{w4}) and 3D  (\ref{w1}) physical
systems with dipole-dipole interactions.   Transfer of polarization disturbances plays a major role
in the information transfer in biological systems. Such
processes do not require particles of the medium to possess
EDM as the dynamics of a system of charged particles
leads to collective polarization.


\begin{thebibliography}{}
\bibitem{2} S. J. Barnett, Phys. Rev. \textbf{6}, 239 (1915).
\bibitem{3}  A. Einstein and W. J. de Haas, Verh. Dtsch. Phys. Ges.
\textbf{17}, 152 (1915).
\bibitem{1} Mamoru Matsuo, Junichi Ieda, Eiji Saitoh, and Sadamichi Maekawa,
Spin-dependent inertial force and spin current in accelerating systems, Phys. Rev. B. \textbf{84}  104410  (2011)
\bibitem{110} Mamoru Matsuo, Jun'ichi Ieda, and Sadamichi Maekawa
Phys. Rev. B \textbf{87}, 115301 (2013)
\bibitem{4} B. Basu, Debashree Chowdhury, Inertial effect on spin orbit coupling and spin transport,
Annals of Phys, \textbf{335}, 47 (2013);
 Spin Transport in non-inertial frame, 	Physica B, \textbf{448}, 155 (2014).
\bibitem{5} Mamoru Matsuo, Junichi Ieda, Eiji Saitoh, Sadamichi Maekawa, Phys. Rev. Lett. \textbf{106}, 076601 (2011)
\bibitem{7}   Edouard B. Sonin, Dynamics of Quantised Vortices in Superfluids, Cambridge University Press, London, (2016)
\bibitem{8}     P. A. Davidson, Turbulence in Rotating, Stratified and Electrically Conducting Fluids,   Cambridge University Press, London, (2013)
\bibitem{9}     Greenspan H., The theory of rotating fluids, Cambridge
University Press, London,     (1968)
\bibitem{10}    Lighthill J., Waves in fluids, Cambridge University Press, London,    (1978)
\bibitem{11}  Cushman-Roisin B.,  Introduction to geophysical fluid dynamics,
Prentice-Hall, Englewood Cliffs,     (1994)
\bibitem{12}       J. Boisson, D. C´ebron, F. Moisy and P. P. Cortet, Earth rotation prevents exact solid body rotation of fluids in the
laboratory, 	EPL, \textbf{98}, 59002 (2012)
\bibitem{13} Pedlosky J.,  Geophysical fluid dynamics, Springer, Heidelberg  (1987)
\bibitem{14}  K. D. Aldridge and L. I. Lumb., Inertial waves identified in the earths fluid
outer core. Nature, \textbf{325}(6103):421 (1987)
\bibitem{15}  H. Bondi and R. A. Lyttleton, On the dynamical theory of the rotation of the
earth. The effect of precession on the motion of the liquid core, Proceedings
of the Cambridge Philosophical Society, \textbf{49}(3):498, (1953)
\bibitem{16} Wen Fu and Dong Lai, Corotational instability, magnetic resonances and global inertial-acoustic
oscillations in magnetized black hole accretion discs, Mon. Not. R. Astron. Soc. \textbf{410}, 399 (2011)
\bibitem{161} Duguet Y., Oscillatory jets and instabilities in a rotating
cylinder, Phys Fluids \textbf{18}, 104104   (2006)
\bibitem{162}  Beardsley RC.,  An experimental study of inertial waves in a
closed cone, Stud Appl Math XLIX(2):187–196  (1970)
\bibitem{17}  L. Messio, C. Morize, M. Rabaud and F. Moisy, Experimental observation using particle image velocimetry of inertial
waves in a rotating fluid,
Exp. in Fluids \textbf{44}, 519 (2008)
\bibitem{18}     C. Morize, F. Moisy, M. Rabaud and J. Sommeria, Dynamics of the anisotropy in decaying rotating turbulence with confinement effects,
 18 - eme Congres Francais de Mecanique, Grenoble (2007).
\bibitem{19}    F. Moisy, C. Morize, M. Rabaud, J. Sommeria, Decay laws, anisotropy and cyclone-anticyclone asymmetry in decaying rotating turbulence
 J. Fluid Mech. \textbf{666}, 5 (2011)
\bibitem{E0} Gregory S. Engel, Tessa R. Calhoun, Elizabeth L. Read, Tae-Kyu Ahn, Tomas
Mancal, Yuan-Chung Cheng, Robert E. Blankenship, and Graham R. Fleming, Evidence
for wavelike energy transfer through quantum coherence in photosynthetic
systems, Nature \textbf{446} (2007), 782.
\bibitem{E00} Hohjai Lee, Yuan-Chung Cheng, and Graham R. Fleming, Coherence Dynamics in
Photosynthesis: Protein Protection of Excitonic Coherence, Science \textbf{316} (2007),
1462.

\bibitem{E6}   Kourosh Afrousheh, Observation of resonant electric
Dipole-Dipole interactions between cold
Rydberg atoms using microwave
spectroscopy, Waterloo, Ontario, Canada (2006)
\bibitem{E7}  H. J. Metcalf, and P. van der Straten, "Laser cooling and trapping", Springer (1999).
\bibitem{E71} M. D. Lukin, M. Fleischhauer, R. Cote, L. M. Duan, D. Jaksch,
J. I. Cirac, and P. Zoller, Phys. Rev. Lett. \textbf{87}, 037901 (2001)
\bibitem{E72} M. S. Safronova, C. J. Williams, and C.W. Clark, Phys.
Rev. A \textbf{67}, 040303 (2003).
\bibitem{E8} J. M. Raimond, G. Vitrant, and S. Haroche, J. Phys. B \textbf{14},
L655 (1981).
\bibitem{E81} I. Mourachko, D. Comparat, F. de Tomasi, A. Fioretti, P. Nosbaum, V. M. Akulin,
and P. Pillet, Many-Body Effects in a Frozen Rydberg Gas, Phys. Rev. Lett. \textbf{80}
(1998), 253.
\bibitem{E9}   K. Singer, M. Reetz-Lamour, T. Amthor, L. G. Marcassa,
and M. Weidemu¨ller, Phys. Rev. Lett. \textbf{93}, 163001 (2004).
\bibitem{E91} E. Vliegen, H. J. Worner, T. P. Softley, and F. Merkt,  Nonhydrogenic Effects in the
Deceleration of Rydberg Atoms in Inhomogeneous Electric Fields, Phys. Rev. Lett.
\textbf{92} (2004), 033005.

\bibitem{E10}   K. Afrousheh, P. Bohlouli-Zanjani, D. Vagale, A. Mugford, M. Fedorov, and J. D. D. Martin, PRL \textbf{93}, 233001 (2004)
    \bibitem{E11} I. Mourachko, D. Comparat, F. de Tomasi, A. Fioretti, P. Nosbaum, V. M. Akulin, and P. Pillet
Phys. Rev. Lett. \textbf{80}, 253 (1998)
 \bibitem{190} L. S. Kuz'menkov and S. G. Maksimov, Theoretical and
Mathematical Physics, \textbf{118}, 227 (1999)
 \bibitem{191} L. S. Kuz'menkov, S. G. Maksimov, and V. V. Fedoseev,
Theoretical and Mathematical Physics, \textbf{126}, 110 (2001)
 \bibitem{192} P. A. Andreev, L. S. Kuzmenkov, M. I. Trukhanova, Phys.Rev.B.,
       A quantum hydrodynamics approach to the formation of new types of waves in polarized two-dimension systems of charged and neutral particles, \textbf{84},   245401  (2011)
 \bibitem{193} P. A. Andreev, M. I. Trukhanova, Russian Physics Journal
\textbf{53}, 1196 (2011)
   \bibitem{230}  Mariya Iv. Trukhanova, Prog. Theor. Exp. Phys., Quantum Hydrodynamics Approach to The Research of Quantum Effects and Vorticity Evolution in Spin Quantum Plasmas,    \textbf{2013}, 111I01 (2013)

 \bibitem{23}   T.  Takabayasi,  Prog. Theor. Phys.   \textbf{14},  283  (1955);

T. Takabayasi, J. P. Vigier, Prog. Theor. Phys.  \textbf{18},  573  (1957);

 T.  Takabayasi, Prog. Theor. Phys. \textbf{70}, 1 (1983);

 T.  Takabayasi,       Prog. Theor. Phys. Suppl. \textbf{4},  1  (1957);
\bibitem{24}   P. R. Holland, P. N. Kyprianidis,  Ann. Inst. Henri Poincare. \textbf{49}, 325 (1988);

P. R. Holland,  J. P. Vigier, Found. Phys. \textbf{18}, 741 (1988);

Peter R. Holland, The Quantum Theory of Motion, Cambridge University Press, (1993);

P. R. Holland, Phys. Lett. A.  \textbf{128}, 9 (1988);
 P. R. Holland,  Phys. Rep. \textbf{169}, 293 (1988)
 \bibitem{25}    H. A. Kramers, Quantentheorie des Electrons and des Strablung (Leipzig, 1938) 259
\end{thebibliography}
\end{document}